\documentclass[twocolumn]{aastex62}
\usepackage{amsmath}	

\usepackage{amsmath}
\usepackage{bm}
\usepackage{mathrsfs}

\accepted{by ApJ}

\shorttitle{Large-scale magnetic field of a thin accretion disk with outflows}
\shortauthors{Li \& Cao}

\begin{document}
\title{The large-scale magnetic field of a thin accretion disk with outflows}

\correspondingauthor{Xinwu Cao}
\email{E-mail: cxw@shao.ac.cn}

\author{Jiawen Li }
\affiliation{Key Laboratory for Research in Galaxies and Cosmology, Shanghai Astronomical Observatory,\\
 Chinese Academy of Sciences, 80 Nandan Road, Shanghai, 200030, China}
\affiliation{University of Chinese Academy of Sciences, 19A Yuquan Road, 100049,
	Beijing, China}

\author{Xinwu Cao}
\affiliation{Shanghai Astronomical Observatory, Chinese Academy of Sciences, 80 Nandan Road, Shanghai, 200030, China}
\affiliation{University of Chinese Academy of Sciences, 19A Yuquan Road, 100049,
	Beijing, China}
\affiliation{ Key Laboratory of Radio Astronomy, Chinese Academy of Sciences,
	210008 Nanjing, China}

\begin{abstract}
The large-scale magnetic field threading an accretion disk plays an
important role in launching jets/outflows. The field may probably
be advected inwards by the plasma in the accretion disk from
the ambient environment (interstellar medium or a companion star).
It has been suggested that the external field can be efficiently dragged inwards in a thin disk with magnetic outflows. We construct a self-consistent global disk-outflow model, in which the large-scale field is formed by the advection of the external field in the disk. The outflows are accelerated by this field co-rotating with the disk, which carry away most angular momentum of the disk and make its structure significantly different from the conventional viscous disk structure.
We find that the magnetic field strength in the inner region of the disk can
be several orders of magnitude higher than the external field
strength for a geometrically thin disk with $H/R \sim 0.1$, if the
ratio of the gas to magnetic pressure
$\beta_{\rm out} \sim  10^2 $ at the outer edge of the disk.
The outflow velocity shows layer-like structure, i.e., it decreases with radius where it is launched. The outflow can be accelerated up to $ \sim 0.2-0.3$c from the inner region of the disk, and the mass loss rate in the
outflows is  $ \sim 10 - 70\%$ of the mass accretion rate at the outer
radius of the disk,  which may account for the
fast outflows observed in some active galactic nuclei (AGNs).
\end{abstract}


\section{Introduction} \label{sec:intro}
It is widely accepted that the large scale magnetic field plays an important role
in acceleration and collimation of the jets and/or outflows \citep*[see reviews of][and the references therein]
{1996astro.ph..2022S,2010LNP...794..233S,2000prpl.conf..759K,
	2007prpl.conf..277P}. The opening configuration magnetic field is a key ingredient in both Blandford-Znajek and  Blandford-Payne mechanisms
\citep*[][]{1977MNRAS.179..433B,1982MNRAS.199..883B}, and the kinetic power of a spinning black
hole or/and the rotating accretion disk is tapped into the jets/outflows by the co-rotating
large scale magnetic field.
The origin of such a magnetic field threading the disk is still not well understood.
It has been suggested that  a dynamo working in an accretion disk can generate the required magnetic field for launching jets/outflows
\citep*[][]{1995ApJ...446..741B,1981MNRAS.195..881P,1981MNRAS.195..897P,1998ApJ...500..703R,1998ApJ...501L.189A}
or the external imposed magnetic fields are transported inwards by the accretion flow \citep*[][]{1974Ap&SS..28...45B,1976Ap&SS..42..401B,1989ASSL..156...99V,1994MNRAS.267..235L,2005ApJ...629..960S}.
Recent numerical simulations done by \cite{2016MNRAS.460.3488S} have shown that a net vertical
 magnetic flux is a necessary  prerequisite for jet formation. This implies that a large scale magnetic field
accelerating jets/outflows may probably be formed by the advection of the external
weak field (e.g., the field threading the interstellar medium or the companion star in X-ray binaries)
\citep*[][]{1974Ap&SS..28...45B,1976Ap&SS..42..401B,1989ASSL..156...99V,1994MNRAS.267..235L,2005ApJ...629..960S}.
A steady magnetic field is formed when the inward advection of the field is
counteracted by the outward field diffusion in the disk \citep*[][]{1994MNRAS.267..235L}.
For the turbulent plasma in an accretion disk, the magnetic diffusivity
$\eta$ is proportional to the turbulent viscosity $\nu$. It is found that
the magnetic Prandtl number, ${\cal P}_{\rm m}$, is almost around unity,  either with the rough estimation or the numerical simulations
\citep*[e.g.,][]{Parker..1979,2003A&A...411..321Y,2009A&A...504..309L,2009A&A...507...19F,2009ApJ...697.1901G}.

The advection of the field in a geometrically thin ($H/R\ll 1$) turbulent
accretion disk is inefficient, because of its small radial velocity
[$v_{R}\propto (H/R)^2$] \citep*[][]{1994MNRAS.267..235L}.
Several models were proposed to handle the issue of the ineffective magnetic field advection
 in conventional turbulent thin disks
\citep*[][]{2005ApJ...629..960S,2009ApJ...701..885L,2012MNRAS.424.2097G,2013MNRAS.430..822G,2013ApJ...765..149C}.
It has been suggested that the magnetic flux threading the out region of the disk can be efficiently transported inwards by the comparatively fast inward moving hot corona around the disk surface, i.e., the ``coronal mechanism"
\citep*[][]{2009ApJ...707..428B}.  The hot gas above the disk can be accreted inward faster than the gas
 inside the thin disk, which, to a certain extent, can solve the difficulty of inefficient magnetic field flux
 transport  in the thin disk
\citep*[][]{2009ApJ...701..885L,2012MNRAS.424.2097G,2013MNRAS.430..822G}.
However, in the context that the jets are produced by the magnetic field transported in the hot corona, the maximum power of the jet almost less than $0.05$ Eddington luminosity, which is too low to account for the strong jets ($ \sim 0.1-1L_{\rm{Edd}} $) observed in some blazars \citep*[][]{2018MNRAS.473.4268C}.
Alternatively, \citet{2013ApJ...765..149C} suggested that the most angular momentum
of the disk can be removed by the magnetically
driven outflows, and therefore the radial velocity of the disk is significantly
higher than that of a conventional turbulent thin disk. In their work, the outflow
solution is obtained with given physical properties of the gas at the disk
surface as the boundary conditions. As most angular momentum of the
disk is carried away by the outflows, the gas in the disk is accreting onto the black hole rapidly. Their local self-consistent accretion
disk-outflow solutions show that even moderately weak fields can cause
sufficient angular momentum loss via magnetic outflows to balance outward
diffusion \citep*[see][for the details]{2013ApJ...765..149C}. Their
calculations are limited to the local case, and the magnetic field configuration/strength is taken as input model parameters in their calculations, however, it is known that
the configuration of the advected field is determined by the whole structure
of the accretion disk (mainly the radial velocity distribution)
\citep*[][]{1994MNRAS.267..235L}. The inclination and strength of the field
at the disk surface are no longer free parameters, which are instead
governed by the whole structure of the disk.

In this work, we consider the global structure of a thin accretion disk
predominantly driven by the magnetic outflows, in which the outflows are calculated
based on the derived global field configuration.
We describe the model and numerical method in Sections \ref{model} and \ref{numerical_method}. 
The results and discussion of the model calculations are given in Section \ref{result}, and the 
last section contains a summary of the work.

\section{Model}\label{model}

For a standard thin disk, the mass accretion is driven by the turbulence
in the disk, which is usually described by $\alpha$-viscosity, and the disk
structure can be derived analytically \citep*[][]{1973A&A....24..337S}.
The disk structure will be altered in the presence of outflows, especially for strong outflows, which can carry both mass and angular momentum from the disk. Magnetically driven outflows from an accretion disk is well described by the Blandford-Payne (BP)  mechanism \citep*[][]{1982MNRAS.199..883B}. As discussed in Section \ref{sec:intro}, a large-scale magnetic field threading the accretion disk is a necessary ingredient of the BP mechanism, which is assumed to be formed by advection of the external weak field. A steady field is achieved when the advection is balanced with the magnetic diffusion in the disk. The magnetic field configuration/strength of the disk can be calculated with the magnetic induction equation when the disk structure (i.e., the radial velocity, density, and temperature as functions of radius) is specified \citep*[][]{1994MNRAS.267..235L}. We note that the magnetic field configuration/strength in the space above the disk is determined by the field strength at the disk surface, which is a global problem, i.e., a change of local disk structure at a certain radius would alter the whole field configuration (not only the field at that radius) \citep*[see][for the details]{1994MNRAS.267..235L}. A fraction of gas at the disk surface can be centrifugally accelerated into outflows by a suitable large-scale magnetic field co-rotating with the disk. Such outflows may carry away a substantial fraction of energy and angular momentum of the disk, which makes the disk significantly different from a standard thin disk.
So, the dynamical structure of the disk is coupled with the outflow solution
\citep[e.g.,][]{1993A&A...276..625F,1995A&A...295..807F,1997A&A...319..340F,2006A&A...447..813F}.

In this work, we construct a steady model of the accretion disk with magnetic driven outflows. We assume a vertical weak magnetic field to be advected by a thin accretion disk to form a strong magnetic field. The outflows are accelerated by the this large-scale magnetic field co-rotating with the disk. Unlike a standard thin disk, the disk considered here is driven both by the turbulence and magnetic outflows. Our model calculations consist of three parts.

1. The structure of the disk is calculated with the magnetic outflows included. Compared with the standard thin disk model, additional terms of the angular momentum and mass loss rate by the outflows should be properly included in the angular momentum and continuity equations of the disk respectively (the details will be described in Section \ref{disk_structure}). The properties of the outflow are to be derived with a suitable outflow model.

2. The properties of the outflow can be derived when the large-scale magnetic field configuration/strength and the rotating velocity of the disk are known with suitable boundary conditions, i.e., the density and temperature of the gas at the bottom of the outflow. The gas in the disk moves into the outflow smoothly, which is a quite complicated problem and has been studied by \citep*[][]{1998ApJ...499..329O,2001ApJ...553..158O}. To avoid this complexity, we assume that the outflow matches to the disk at the disk surface with a scale-height $H$ (hereafter we refer to the place $z\sim H$ as either the bottom of the outflow or the disk surface). The gas in the outflow is driven from the disk surface, and its temperature should be the same as the disk surface temperature. The gas density of the disk varies significantly in vertical direction, and the physics of the transition region from the disk to the outflow is very complicated and quite uncertain \citep*[see][for the detailed discussion]{2001ApJ...553..158O}. To avoid such complexity, we use a parameter $\beta_{\rm s}$ (the ratio of gas to magnetic pressure) to estimate the gas density at the disk surface $H$. In order to magnetically accelerate the gas at the disk surface into the outflow, the magnetic pressure should be greater than the gas pressure at the bottom of the outflow/the disk surface, i.e., $\beta_{\rm s}\la1$ is required. The detailed calculations of an outflow from the disk for given magnetic field configuration/strength with boundary conditions at the disk surface will be described in Section \ref{outflow_solution}.

3. The poloidal field advection/diffusion balance in the accretion disk is described by the magnetic induction equation. We assume a vertical weak external field to be dragged inwards by the disk extending from $R_{\rm in}$ to $R_{\rm out}$. Above the disk, a potential field is assumed, which is a good approximation for a tenuous outflow, and is widely adopted in most previous modeling of magnetically driven outflows. The whole magnetic field configuration/strength of the disk are  available by solving the induction equation as done by \citet{1994MNRAS.267..235L} (see Section \ref{field_config} for the details).

We note that three parts in our model calculations are not separable. For example, we need to know the outflow properties when calculating the disk structure, while the calculations of the outflow requires field configuration/strength of the disk that is derived from the disk structure. So, the calculations of the model are carried out by numerical iterations.

\subsection{Structure of an accretion disk with magnetically driven outflows}
\label{disk_structure}

We consider an accretion disk with outflows centrifugally accelerated by the magnetic field lines co-rotating with the disk.
A fraction of the gas is driven from the disk surface, which may remove most angular momentum of the disk, and an extra torque
caused by the outflows is exerted on the disk. In this case, the accretion
is driven both by the magnetic torque and the turbulence in the disk.
As usual, in cylindrical coordinate, we use $ \dot{M}_{\rm{acc}} (R)$ to describe the mass inflow rate in the disk,
\begin{equation}
\dot{ M}_{\rm{acc}} (R) = -2\pi R\Sigma v_{R},\label{mdot}
\end{equation}
where $\Sigma \simeq 2\rho H$ is the disk surface density, $ \rho $, $ H $ are the mean
disk density and the scale height of the disk, respectively. We note that the mass accretion rate $ \dot{ M}_{\rm acc} $
is a function of $ R $ due to the presence of outflows, which is related to the mass loss rate,  $ \dot{m}_{\rm w} $,
from the unit area (one surface) of the disk by the continue equation,
\begin{equation}
\frac{d \dot{M}_{\rm{acc}} }{d R} = 4\pi R\dot{m}_{\rm w}.\label{dM_acc-dR}
\end{equation}

The momentum equation of the disk with magnetic outflows reads
\begin{equation}\label{angular_eq}
\frac{d}{d R}\left( 2\pi R \Sigma v _{ R} R^2 \Omega \right) = \frac{d}{d R}
\left( 2 \pi R \nu \Sigma R^2 \frac{d \Omega}{d R} \right) + 2 \pi R T_{\rm m} ,
\end{equation}
of which the first term in the right side describes the angular momentum transfer caused by turbulence, and the second term is due to the magnetic outflows.
The radial velocity, $v _R$, of an accretion disk with magnetic outflows can be calculated by integrating Equation \eqref{angular_eq}, which reads
\begin{equation}\label{v_R}
\begin{aligned}
v _{ R}=&   v_{R,  \rm{vis}} + v_{R,\rm m}\\
=&-\frac{3\nu}{2R} - \frac{T_{\rm m}}{\Sigma}\left[ \frac{\partial}{\partial R}\left( R^2 \Omega\right)  \right] ^{-1}\\
=&-\frac{3\nu}{2R} - \frac{2T_{\rm m}}{\Sigma R \Omega},
\end{aligned}
\end{equation}
where the $ \alpha $-viscosity $ \nu =\alpha c_{\rm {s,c} } H$ is adopted.
The approximation $d \Omega /dR \simeq - 3\Omega /2R$ is used in deriving Equation \eqref{v_R}.

For a standard thin accretion disk, its vertical structure is derived with the momentum equation by assuming the gas pressure gradient is in equilibrium with the vertical component of the gravity of the black hole \citep*[][]{1973A&A....24..337S}. In the presence of a large-scale magnetic field, the disk structure is altered significantly especially in the strong field case, because the curved field lines exert forces on the gas of the disk both in vertical and radial directions \citep*[e.g.,][]{1998ApJ...499..329O,2001ApJ...553..158O,2002A&A...385..289C}. One can, in principle, solve the momentum equations in radial and vertical direction under some suitable boundary conditions to derive the magnetic field configuration \citep*[e.g.,][]{2001ApJ...553..158O}. This is too complicated to be included in present work. A general calculation of the field configuration is described by Kippenhahn-Schl{\"u}ter (KS) model \citep*[][]{1957ZA.....43...36K}, which was established for calculating the magnetic field configuration within the solar filaments. It describes the filaments supported by the a magnetic field, which can reproduce the observations well. The KS model is valid for an isothermal plasma sheet suspended against gravity by a magnetic field, which is similar to an accretion disk vertically compressed by a magnetic field. For an isothermal accretion disk considered in this work, the KS model is applicable. Considering the cumbersome theoretical form in their model, an approximated fitting formula for KS model is given by \cite{2002A&A...385..289C}, which can reproduce the results of the KS model at a fairly good accuracy. It reads
\begin{equation}
R-R_0 = \frac{H}{\kappa_0 \eta^2  _{ i}}\left( 1 - \eta^2 _ i + \eta^2 _ i z^2H^{-2} \right)^{1/2} - \frac{H}{\kappa_0 \eta^2 _ i}\left( 1-\eta^2 _ i  \right)^{1/2}, \label{b_shape}
\end{equation}
 where $R_0$ is the radial distance of the magnetic field line footpoint, $ \eta _  i = \rm{tanh}(1) $, $ \kappa_0 = B_{\rm z}/B_{R,\rm s}$ is the inclination of the magnetic field line with respect to horizontal at the disk surface $z=H$, and $B_{R,{\rm s}}$ is the radial component of magnetic field at disk surface.

The disk is compressed in the vertical direction both by the magnetic stress (which acts like a negative pressure) and the vertical component of the gravity. With the field line shape described by Equation (\ref{b_shape}), one can solve the vertical momentum equation of an isothermal disk,
\begin{equation}
c_{\rm s,s}^2{\frac {d\rho(z)}{dz}}=-\rho(z)\Omega_{\rm K}^2z-{\frac {B_R(z)}{4\pi}}{\frac {dB_R(z)}{dz}},
\label{dpdz}
\end{equation}
numerically assuming the gas to be in equilibrium vertically, and the scale-height of the disk compressed by the magnetic field is available. A fitting formula suggested by \cite{2002A&A...385..289C}
\begin{equation}\label{H/R_cs02}
\begin{aligned}
\frac{H}{R} =& \frac{1}{2}\left[ \frac{4 c^2_{\rm s, \rm c}}{R^2  \Omega ^2_{\rm k}} + \frac{\left( \Omega ^2_{\rm k} - \Omega ^2\right)^2 }
{4\left( 1 - {\rm e}^{-1/2} \right)^2 \kappa^2 _0 \Omega^4_{\rm k} } \right]^{1/2}\\
&- \frac{ \Omega ^2_{\rm k} - \Omega ^2 }{4\left( 1 - {\rm e}^{-1/2} \right) \kappa _0 \Omega^2_{\rm k} },
\end{aligned}
\end{equation}
which can reproduce the numerical results quite well.

The disk is rotating in a sub-Keplerian velocity, because a radial magnetic force is exerted on the gas in the disk against the gravity, i.e.,
\begin{equation}\label{diff_omega}
R\left( \Omega ^2 _{\rm k} - \Omega ^2 \right) = \frac{B_{z} B_{R,\rm s}}{2\pi \Sigma},
\end{equation}
which yields
\begin{equation}\label{ome/ome_k}
f_{\Omega} \equiv \frac{\Omega}{\Omega_{\rm k}} = \left[ 1 - \frac{2R\kappa_ 0}{\beta \left( 1 + \kappa^2 _0 \right)H } \frac{c^2_{\rm{s,c}}}{R^2 \Omega^2 _{\rm k}}\right] ^{1/2}.
\end{equation}
Using Equations \eqref{H/R_cs02} and \eqref{diff_omega}, we obtain
\begin{equation}\label{H_over_R}
\frac{H}{R} = \frac{c_{\rm{s,c}}}{R \Omega_{\rm k}}\left[ 1 - \frac{1}{\left( 1 - \rm{e}^{-1/2}  \right)  \left(  1+\kappa^2 _0   \right) \beta  }  \right]^{1/2},
\end{equation}
where $\Sigma=2\rho H$, $\kappa_0 =B_{z}/ B_{R, \rm s}$, and $ \beta $ is the ratio of the gas pressure to the magnetic pressure at mid-plane of the disk defined as
\begin{equation}\label{beta_p}
\beta  \equiv \frac{P_{\rm {gas}}}{P_{\rm{mag}}} ={\frac {8\pi P_{\rm {gas}}} {B^2}}.
\end{equation}

The structure of the disk with magnetic outflows is available by solving a set of equations described above, provided the magnetic strength/configuration is known. We note that the magnetic torque $T_{\rm m}$ and mass loss rate $m_{\rm w}$ in outflows can be calculated with a suitable outflow model for a given magnetic field, which will be described in Sect. \ref{outflow_solution}.

\subsection{Magnetic outflows from the disk}\label{outflow_solution}

In principle, the solution of the magnetically driven outflow is available by solving a set of MHD equations when the field strength and configuration are specified with suitable boundary conditions at the bottom of the outflow (i.e., the surface of the disk) \citep*[e.g.,][]{1994A&A...287...80C,2014ApJ...783...51C}, which is an extension of the Weber-Davis model in the disk case (i.e., an axis-symmetrical rotating system). The initial Weber-Davis model was developed to explore the winds accelerated by an ideal split monopole magnetic field threading a rotating star \citep*[][]{1967ApJ...148..217W}, which has a spherical symmetry. In the limit of the cold gas, i.e., the gas pressure is negligible in a magnetically driven outflow, a simple description of the outflow acceleration is available in terms of the field strength and mass loss rate
\citep*[][]{1996astro.ph..2022S,Mestel2012}. It was found that the cold Weber-Davis model applied for outflows magnetically driven from an accretion disk can reproduce the results of the numerical simulations fairly well \citep*[][]{2005ApJ...630..945A}.
In this work, the outflow is calculated by using the approach suggested by \cite*{2013ApJ...765..149C} in the frame of the cold Weber-Davis model, which is briefly summarized as follows.

The dimensionless mass loading parameter $ \mu $ is defined as \citep*[see][for the details]{1996astro.ph..2022S,2013ApJ...765..149C}
\begin{equation}
\mu = \frac{4\pi \rho _{\rm w}v_{\rm w}\Omega R}{B^2_{\rm p}}
 ={\frac{4\pi \Omega R \dot{m}_{\rm w}}{B_{z}B_{\rm p}}}, \label{mass_load_para}
\end{equation}
where $ \rho _{\rm w}v_{\rm w} =  \dot{m}_{\rm w} B_{\rm p}/B_{z}$ is used, which is the mass flux parallel to the field
line, $ \Omega (R) $ is the angular velocity of the field line corotating with the disk at its footpoint,
 and $ B_{\rm p} =( B^2 _{z} + B^2 _{R} )^{1/2}  $ is the poloidal component of the magnetic field. For a given magnetic field line anchored in the disk at radius $R$ rotating with the disk,
the dimensionless mass loading parameter $\mu$ indicates how much mass is loaded along the field line compared with the mass accretion rate in the disk. It also determines the terminal speed of the gas accelerated by this magnetic field. The gas in the outflow will be centrifugally accelerated up to Alfv\'en point, at which the flow speed becomes comparable to the Alfv\'en speed. This is a good approximation verified by numerical magnetic outflow model calculations \citep*[][]{1996astro.ph..2022S}. Beyond the Alfv\'en point, the kinetic energy of the gas in the outflow $\rho v_{\rm w}^2/2$ will dominate over the magnetic energy $B^2/8\pi$, and therefore further acceleration of the gas by the magnetic field is rather inefficient. The total magnetic torque $ T_{\rm m} $ per unit area on the disk (including both sides) can be calculated from the mass loss rate in the outflow
and the  Alfv\'en radius
\begin{equation}\label{T_m}
T_{\rm m} \simeq 2\dot{m}_{\rm w}R^2 _{\rm A} \Omega ,
\end{equation}
where $ R_{\rm A} $ is the Alfv\'en radius of the outflows \citep*[][]{1994MNRAS.267..235L}. The solution of cold Weber-Davis model yields
\citep[see][for detailed discussion]{1996astro.ph..2022S,Mestel2012}:
\begin{equation}\label{Alfven_Radial}
R_{\rm A} = R \left[ \frac{3}{2} \left( 1 + \mu^{-2/3} \right)  \right]^{1/2},
\end{equation}
where $ R $ is the radial distance of the magnetic field line footpoint from the central black hole.

Using Equations \eqref{mass_load_para}  and  \eqref{T_m} we obtain
\begin{equation}\label{T_m_nu}
T_{\rm m} = \frac{3}{4\pi}RB_{z} B_{\rm p}\mu \left( 1+\mu^{-2/3} \right).
\end{equation}
In most of our model calculations, the accretion disk is predominantly driven by magnetic outflows, i.e., $v_{R,{\rm m}}\gg v_{R,{\rm vis}}$ in Equation (\ref{v_R}). In this case,
\begin{equation}\label{v_R2}
v _{ R}\simeq v_{R,\rm m}
=-\frac{2T_{\rm m}}{\Sigma R \Omega},
\end{equation}
is a good approximation. We substitute Equations (\ref{mass_load_para}), (\ref{T_m_nu}) and (\ref{v_R2}) into Equation (\ref{dM_acc-dR}), we have
\begin{equation}
{\frac {d\ln\dot{M}_{\rm acc}}{d\ln R}}={1 \over 3}(1+\mu^{-2/3})^{-1}, \label{mdot_modtw}
\end{equation}
where Equation (\ref{mdot}) is used. It indicates that the mass load parameter $\mu$ describes the relative importance of the mass loss rate in outflows with the mass accretion rate in the disk.
Substituting Equation \eqref{T_m_nu} into
Equation \eqref{v_R}, we have
\begin{equation}\label{vr1}
v_{R} = -\frac{3\nu}{2R} -\frac{6  c^2 _{\rm{s,c}} B_{z} }{\beta H \Omega B_{\rm p}}\mu \left( 1+\mu^{-2/3} \right),
\end{equation}
where $ c_{\rm s, \rm c}  $ is the sound speed at the mid-plane of the disk.

We note that the dynamics of the outflow driven from the disk is described analytically, provided the field strength at the disk surface and the mass load parameter $\mu$ are specified. However, the dimensionless parameter $\mu$ can be derived only if the mass loss rate $\dot{m}_{\rm w}$ in the outflow along the field line is known (see Equation \ref{mass_load_para}).

The Bernoulli equation of the gas along a field line in an isothermal outflow is
\begin{equation}
{\frac {1}{2}}v_{\rm w}^2+c_{\rm s,s}^2\ln\rho_{\rm w}+\Psi_{\rm eff}={\rm const},
\label{bernoulli}
\end{equation}
where $c_{\rm s,s}$ is the sound speed \citep*[e.g.,][]{1994A&A...287...80C,1998ApJ...499..329O}. The effective potential is
\begin{equation}
\Psi _{\rm eff}(R,z) = - {\frac {GM_{\rm BH}}{( R^2 + z^2)^{1/2}}} - {\frac{1}{2}}R^2 \Omega ^2, \label{psi}
\end{equation}
where $\Omega$ is the rotation rate of the footpoint at disk surface, $M_{\rm BH}$ is the central black hole mass. Differentiating Equation (\ref{bernoulli}) along the field line, we obtain
\begin{equation}
(c_{\rm s,s}^2-v_{\rm w}^2){\frac {d\ln\rho_{\rm w}}{dl}}+{\frac {d\Psi_{\rm eff}}{dl}}=0.
\label{dif_bernoulli}
\end{equation}
At the sonic point in the outflow, $v_{\rm w}=c_{\rm s,s}$, $d\Psi_{\rm eff}/dl=0$, which implies that the location of the sonic point corresponds to the maximum of the effective potential, i.e., $\Psi_{\rm eff,s}=\Psi_{\rm eff,max}$. Thus, the location of the sonic point in the outflow can be easily derived when the magnetic field configuration and the disk rotational rate are specified. Using the Bernoulli equation, one can derive the relation of the physical quantities at the bottom of the outflow with those at the sonic point as
\begin{equation}
{1\over 2}\left[1-\left({\frac {v_{\rm w,0}}{c_{\rm s,s}}}\right)^2\right]+\ln{\frac {v_{\rm w,0}}{c_{\rm s,s}}}+{\frac {1}{c_{\rm s,s}^2}}(\Psi_{\rm eff,s}-\Psi_{\rm eff,0})=0,
\label{rho_s}
\end{equation}
where the subscript ``0" refers to the quantities at the bottom of the outflow, and the mass conservation $\rho_{\rm w,s}c_{\rm s,s}=\rho_{\rm w,0}v_{\rm w,0}$ is used. As $v_{\rm w,0}/c_{\rm s,s}\ll 1$, we have
\begin{equation}\label{rho_s2}
\begin{aligned}
\rho_{\rm w,s}&\simeq \rho_{\rm w,0}\exp\left(-{\frac {\Psi _{\rm{eff,s}}-\Psi _{\rm eff,0}}{c^2_{\rm{s,s}}}}-{1\over 2}\right)\\
&\simeq\exp\left(-{\frac {\Psi _{\rm {eff,s}}-\Psi _{\rm eff,0}}{c^2_{\rm s,s}}}\right),
\end{aligned}
\end{equation}
where $\Psi_{\rm eff,s}-\Psi _{\rm eff,0}\gg c_{\rm s,s}^2$ is always satisfied for a magnetically driven outflow (i.e., the gas pressure is negligible in the outflow acceleration).

The mass loss rate in the outflow from unit area of an accretion disk is available,
\begin{equation}\label{dot-m_w}
\dot{m}_{\rm w} \simeq \frac{B_{z}}{B_{\rm p}} \rho _{\rm w,0} c_{\rm{s}}\rm{exp}\left( -\frac{\Psi _{\rm{eff,max}} - \Psi _{\rm{eff,0}} }{c^2_{\rm{s,s}}}  \right),
\end{equation}
where $\Psi _{\rm eff,s}=\Psi _{\rm eff,max}$ because the sonic point is always located at the maximum of the effective potential  (see Equation \ref{dif_bernoulli}).

It is obvious that shape of the magnetic field line is required to calculate the effective potential along the field line, and then to derive the maximal value of the potential. The position of the sonic point $ z_{\rm s} $, and the maximal effective potential $ \Psi_{\rm{eff,s}} $, can be calculated with this field configuration when $\kappa_0$, and the angular velocity $ \Omega $ of the disk, are known.

The temperature of the gas is crucial in calculating the mass loss rate in the outflows,  which is assumed to  be the
same as the  disk surface temperature $ T_{\rm s} $.
For a thin accretion disk, the temperatures at the disk surface and the disk mid-plane are related by
$\sigma T^4_{\rm s} = (4/3\tau)\sigma T^4_{\rm c}$, this leads to
\begin{equation}\label{c_sc-c_ss}
c_{\rm{s,s}} = \left( \frac{4}{3\tau} \right) ^{1/8} c_{\rm{s,c}},
\end{equation}
where $ \tau $ is the optical depth in vertical direction of the disk.

 In order to calculate mass loss rate per unit area of the disk,
$ \dot{m}_{\rm w} $, with Equation \eqref{dot-m_w}, we need to know
the density at the base of the outflow. The density of the gas at the disk surface is quite uncertain,
however, we can estimate an upper limit on the gas density at the bottom of the outflow.
 Within the  Alfv\'en radius the magnetic field is strong enough to enforce the gas corotating with
  the field line. The magnetic pressure is assumed to be dominant over the gas pressure at the bottom of the
  outflow, i.e., $ B^2_{\rm p}/8\pi \gtrsim \rho_{\rm w,0} c^2_{\rm{s,s}}$, so the gas will move along the field lines. Otherwise, the field will be frozen in and move with the gas if $\beta_{\rm s}>1$. This implies that outflow can be launched from the disk surface only if the field strength is sufficiently strong. Let $ \beta_{\rm s} $ be the plasma beta
parameter at the base of the flow $ (\beta_{\rm s} \lesssim 1 )$, we have
\begin{equation}\label{rho_0-c_ss}
\rho_{\rm w, 0} c_{\rm{s,s}}  =  \beta_{\rm s}\left(  \frac{3\tau}{4}\right)^{1/8}  \frac{B^2_{\rm p}}{8\pi c_{\rm{s,c} }},
\end{equation}
where Equation \eqref{c_sc-c_ss} is used.

Using Equations \eqref{mass_load_para}, \eqref{dot-m_w}, and \eqref{rho_0-c_ss} we find
\begin{equation}\label{mass_load_para_solve-2}
\dot{m}_{\rm w}=\frac{\beta_{\rm s }B_{\rm z}B_{\rm p}}{8 \pi c_{\rm{s,c}}}
 \left( \frac{3\tau}{4} \right) ^ \frac{1}{8}  {\rm exp} \left( -\frac{\Psi _{\rm{eff,s}} -
 	\Psi _{\rm{eff,0}} }{c^2_{\rm{s,s}}}  \right),
\end{equation}
where $ z_{\rm s} $ is the vertical position of the sonic point.
The total mass loss rate in the outflows is
\begin{equation}\label{dot_M_w}
\dot{M}_{\rm w} =  \int\limits_{R_{\rm{in}}}^{R_{\rm{out}}} 4\pi R \dot{m}_{\rm w}d R
\end{equation}
where $ R_{\rm{in}} $ and $ R_{\rm{out}} $ are the inner and outer edge of the disk respectively.

For given boundary conditions of the outflow at the disk surface (i.e., temperature and density of the gas), the properties of the outflow is available with the calculations described in this section with a specified large-scale magnetic field. The magnetic torque $T_{\rm m}$ and the mass loss rate of the outflows derived here are then adopted in the calculations of the disk structure (see Section \ref{disk_structure}).

The terminal speed of the outflows can be derived by solving the Bernoulli equation, i.e., the equation of motion of the flows along the magnetic field lines, which describe that the total energy of the kinetic, thermal, gravitational, and a so-called `centrifugal energy'
is constant along the field line. We use the result derived in \cite{1996astro.ph..2022S} [Equation (74) in Sect. 7] to calculate the terminal speed of the outflows, i.e.,
\begin{equation}\label{termial-speed}
v_\infty = R_0\Omega \mu ^{-1/3},
\end{equation}
where $ R_0 \Omega $ is the rotational velocity of the magnetic field foot-point. The terminal speed $ v_{\infty} $ of the outflow is directly related to the mass load parameter $\mu$, which indicates a fast outflow related to a low mass loading (low-$ \mu $), while a slow outflow relate to a high mass loading (high-$ \mu $).

\subsection{Magnetic field configuration of the disk with outflows}
\label{field_config}
As described in Section \ref{outflow_solution}, the mass loss rate of the outflows can be calculated
when the field configuration and strength are specified. With the derived mass loss rate, the torque
exerted on the disk caused by the magnetic outflows is available (Eq. 4), and the disk structure can be calculated by integrating the angular momentum equation of the disk.

The evolution of poloidal magnetic field threading an accretion disk is described by the induction equation \citep*[see Equation 10 in][]{1994MNRAS.267..235L},
\begin{equation}\label{induc_psi}
\frac{\partial }{\partial t} \left( R \psi \right) = -v_{R}
\frac{\partial }{\partial R}
\left( R \psi \right) - \frac{4 \pi \eta}{c}R J_ \phi,
\end{equation}
where $v_{R}$ is the radial accretion velocity, $\eta$ is the magnetic
diffusivity, and $J_ \phi$ is azimuthal current density of the disk. $ R \psi $ is proportional
to the magnetic flux through the disk within radius $ R $, which is a constant along the magnetic field line.
Integrating Equation (\ref{induc_psi}) over the vertical direction, we have
\begin{equation}
\label{induc_psi_average}
\frac{\partial }{\partial t} \left[ R \psi (R,0) \right] = -v_{R}
\frac{\partial }{\partial R} \left[ R \psi (R,0) \right] - \frac{4 \pi \eta}{c}
\frac{R}{2H} J_ \phi ^{\rm s},
\end{equation}
where
\begin{equation*}
J_ \phi ^{\rm s} \equiv \int\limits_{-H}^{H} J_ \phi (R,z) dz.
\end{equation*}
The surface current density of the disk $ J_ \phi ^{\rm s} $ is related to the magnetic potential $ \psi $
via the Biot-Savart law,
\begin{equation}\label{psi_d}
\begin{aligned}
\psi _{\rm d} (R,0) =& \psi (R,0) - \psi _ \infty (R,0) \\
=& \frac{1}{c}\int\limits_{R_{\rm in}}^{R_{\rm out}}\int\limits_{0}^{2\pi}
\frac{J_\phi ^{\rm s} (R')cos\phi'  d\phi' R' dR' }{(R^2 + R'^2 -
	2RR'cos\phi')^\frac{1}{2}},
\end{aligned}
\end{equation}
where $ \psi _{\rm d} (R,0) $ due to currents inside the accretion
disk. Assuming a uniform external vertical magnetic field $B_{\rm{ext}}$, and therefore
\begin{equation*}
\psi _ \infty (R,0) = \frac{1}{2} B_{\rm{ext}} R.
\end{equation*}
For a steady accretion disk-outflow system, i.e., $\partial \psi/\partial t=0$, the advection/diffusion
of the steady large scale magnetic field in the disk is described by
\begin{equation}\label{induc_psi_average_stab}
\begin{aligned}
-{\frac {\partial }{\partial R}}[R\psi_ {\rm d} (R,0)] - {\frac {2\pi}{c}}{
	\frac {\alpha c_ {\rm {s,c}} R}{v_{R}}}{\cal P}_{\rm m} J_{\phi}^ {\rm s} (R)=B_{\rm{ext}} {R},
\end{aligned}
\end{equation}
where $ {\cal P} _{\rm m} \equiv \eta/\nu $, is the magnetic Prandtl number. Differentiating Equation \eqref{psi_d}, we have
\begin{equation}\label{d_psi_d}
\begin{aligned}
&{\frac {\partial}{\partial R}}[R\psi_ {\rm d}(R,0)]=\\
&\frac{1}{c}\int\limits_{R_{\rm in}}^{R_{\rm out}}\int\limits_{0}^{2\pi}
\frac{(R'^2 - RR'cos\phi')J_\phi
	^{\rm s} (R')cos\phi'  d\phi' R'dR'}{(R^2 + R'^2
	- 2RR'cos\phi')^\frac{3}{2}}.
\end{aligned}
\end{equation}
Substituting this equation into Equation \eqref{induc_psi_average_stab}, we obtain a group of linear equations (the subscript $ _{i,j} $ is labeled for the variables at radius $ R_i $, $R_j  $)
\begin{equation}\label{linear_eq}
-\sum\limits_{j=1}^{n}P_{ij}{J_{\phi}^{\rm s }(R_j)}\Delta{R_j}
-{\frac {2\pi}{c}}{\frac {\alpha c_ {\rm {s,c}}(R_i)R_i}{v_{R}(R_i)}}
{\cal P}_{\rm m}J_{\phi}^{\rm s} (R_i)=B_{\rm{ext}}{R_i},
\end{equation}
where
\begin{equation*}
P_{ij}= {\frac{1}{c}}\int\limits_{0}^{2\pi}{\frac {\left( R_j -
		R_i \cos\phi' \right) R_j ^2}{\left( {R_i}^2+{R_j}^2-2R_i{R_j}
		\cos{\phi '}\right) ^{\frac{3}{2}}}}\cos \phi'{d}\phi ',
\end{equation*}
$J_{\phi}^{\rm s }(R_j) $ is the surface current density of the circular ring
at radius $R_j$ with a width of $\Delta R_j$. The radial accretion velocity caused both by the angular
momentum transport in the outflows and turbulence is determined by equation \eqref{vr1},
\begin{equation}\label{vr2}
v_{R}(R_i) = -\frac{3}{2}  \frac{\alpha c_{\rm s,c} (R_i) H_i }{R_i} - \frac{6 \kappa_{0,i} \mu_{i} \left( 1+ \mu _{i}^{-2/3} \right)c_{\rm s,c} (R_i) }{ \beta_{i} f_{\Omega, i} \left( 1+ \kappa^2_{0,i} \right)^{1/2} }.
\end{equation}

Solving the linear system (Eq. \ref{linear_eq}) with given disk structure, i.e., the radial velocity and the disk thickness, etc. We can obtain the distribution of the
surface current density, and then the magnetic field potential. The magnetic potential in the space above the disk is
\begin{equation}
  \psi_{\rm d} (R,z)=\frac{1}{c}\int\limits_{R_{\rm in}}^{R_{\rm out}}\int\limits_{0}^{2\pi}
 \frac{J_\phi^{\rm s}(R')cos\phi'  d\phi' R' dR'}{\left[R^2 + R'^2 +
 z^2 - 2RR'cos\phi'\right]^\frac{1}{2}}.\label{psi_d_2}
 \end{equation}
The poloidal magnetic field configuration
can be calculated, i.e.,
\begin{equation}\label{br}
B_{R} = -\frac{\partial \psi}{\partial z},
\end{equation}
and
\begin{equation}\label{bz}
B_z = \frac{1}{R} \frac{\partial}{\partial R}\left( R \psi \right).
\end{equation}

\subsection{Comparison with previous works}\label{comparison}

There are some similar previous works closely related with our present work. \citet{1994A&A...287...80C} explored the dynamical properties of outflows accelerated by the magnetic field threading the rotating disk, in which a simplified magnetic field configuration in an analytical form is adopted to mimic a realistic field threading the disk. The temperature and density of the gas at the bottom of the outflow are taken as input model parameters. Their calculations focus on the dynamics of the gas accelerated along the field lines varying with different inclination angles. It is found that the magnetic field configuration plays a crucial role in launching outflows \citep[][]{1994A&A...287...80C}, though the origin of such large-scale field is not included in their work. Such a large-scale magnetic field may probably be formed through advection of external weak field threading ambient gas by the accretion disk. \citet{1994MNRAS.267..235L} calculated how the weak external vertical magnetic field is dragged inwards by a standard thin accretion disk, in which a steady field is formed with advection and diffusion balanced. It was found that the advection of the field in a thin disk is quite inefficient due to its small radial velocity, which implies the difficulty for launching an outflow from a thin disk. Their calculations focus on the field configuration assuming the structure of the disk is not altered by the field, and the gas acceleration along the field line has not been considered in their model.

The magnetic outflow matches the disk at some place very close to the disk surface. As the vertical disk structure is complicated and it may also affect by the outflow, the vertical transition region from the disk to outflow has been studied in previous works \citep*[][]{1998ApJ...499..329O,2001ApJ...553..158O}. In their model calculations, the disk structure altered by the magnetic field is considered locally (for a certain radius), and the removal angular momentum from the disk by the outflow is not included. They also concluded that launching an outflow magnetically from a thin disk is difficult. In order to solve the difficulty of launching outflows from a thin accretion disk, \citet{2013ApJ...765..149C} suggested that the radial velocity is substantially increased if most angular momentum of the gas in a thin disk is removed through outflows, and therefore the weak external field can be advected inwards to form a strong field near the black hole. Their calculations were carried out for the outflow along a single field line with specified field strength and field line inclination at the disk surface as input parameters, however, the field strngth/configuration are functions of radius, which are mainly determined by the whole structure of the disk \citep*[e.g.,][]{1994MNRAS.267..235L}. Thus, their results show that field advection in the thin disk with outflows could be very efficient locally, however, it is not sure if such mechanism works for a realistic disk (i.e., if it works at all radii of the disk). In principle, the external weak field advected by an accretion disk predominantly driven by magnetic outflows is an accretion disk-outflow coupled global problem. In this work, we present a self-consistent accretion disk-outflow solution, in which we incorporate the calculations of \citet{2013ApJ...765..149C} for local outflow structure with the global field calculated with \citet{1994MNRAS.267..235L}'s method. We have not only derived the global configuration of the field advected by the disk with outflows, but also the mass loss rate and terminal speed of the outflow together with the disk structure as functions of radius. We note that a series of works have been carried out on the disk-outflow system, in which the removal of angular momentum of the gas in the disk by the magnetic outflows is properly considered. However, a rather simplified self-similar disk-outflow solution is derived in their calculations \citep*[][]{1993A&A...276..625F,1995A&A...295..807F,1997A&A...319..340F,2006A&A...447..813F}.

\section{Numerical method}\label{numerical_method}

The model of an accretion disk with outflows is described in Section \ref{model}, which can be calculated numerically when the model parameters, $\alpha$, $\beta_{\rm out}$, ${\cal P}_ {\rm m}$, $ \Theta = c^2_{\rm s,c}/(R^2\Omega^2_{\rm k})$ (the dimensionless temperature of the disk), and $ \tau $(the optical depth of the disk) are specified. As described in Section \ref{model}, the model calculations consist of three parts: 1. the structure of the disk with outflows; 2. the outflow solution; 3. the field configuration of the disk. The calculations of any part need the results of other two parts as input ingredients, which make the numerical calculations quite complicated. The numerical method adopted in this work is briefly described below.

1. The model calculations are started with a certain (tentative) initial radial velocity distribution $v_{R}(R)$ of the disk. The magnetic field advection/diffusion in the disk is described by the induction equation, which is an integral-differential equation. It can be transformed to a set of linear algebraic equations, and then be solved numerically \citep*[][]{1994MNRAS.267..235L}. We consider an accretion disk extending from $R_{\rm in}$ to $R_{\rm out}$, which is divided into $n$ grids radially with equal width in logarithmic space ($n=200$ is adopted in our calculations). With this tentative $v_{R}(R_i)$, the surface current density $J_ \phi ^{\rm s}(R_i)~(i=1,n)$ is available by solving a set of $n$ linear algebraic equations numerically (see Equation \ref{linear_eq}). The magnetic potential $R_i\psi_i$ is calculated from derived the surface current density $J_ \phi ^{\rm s}(R_i)$ with Equation (\ref{psi_d_2}). Finally, derivatives of $R\psi$ give the poloidal magnetic field configuration/strength (see Equations \eqref{br} and \eqref{bz}).

2. In the presence of this magnetic field, the disk is rotating at a sub-Keplerian velocity and compressed vertically. The rotating rate $\Omega$ and scale-height $H/R$ of the disk are calculated with Equations (\ref{ome/ome_k}) and (\ref{H_over_R}), and the effective potential $\Psi_{\rm eff}$ along the field line can easily be calculated. Now we can calculate the mass loss rate in outflows from unit area of the disk surface $\dot{m}_{\rm w}$ as a function of radius using Equation (\ref{dot_M_w}), with which the mass loading parameter $\mu$ and magnetic torque $T_{\rm m}$ are derived with Equations (\ref{mass_load_para}) and (\ref{T_m_nu}) respectively.

3. The mass accretion rate of the disk with outflows is no longer constant radially. With derived mass loss rate $\dot{m}_{\rm w}$, one can obtain the mass accretion rate as a function of radius, and then the gas density of the disk is derived. The radial velocity of the accretion disk with outflows $v_R(R)$ is obtained by calculating with Equation (\ref{vr1}), in which the magnetic field strength/configuration obtained in step 1 are used.

The calculation is started with a tentative $v_R(R)$ distribution, and then it arrives a new radial velocity distribution for the disk. A self-consistent solution of an accretion disk-outflow system is achieved, only these two velocity distributions converge. We find that such a solution is indeed available after several iterations of step 1-3 described above.

\begin{figure}
	\includegraphics[width=\columnwidth]{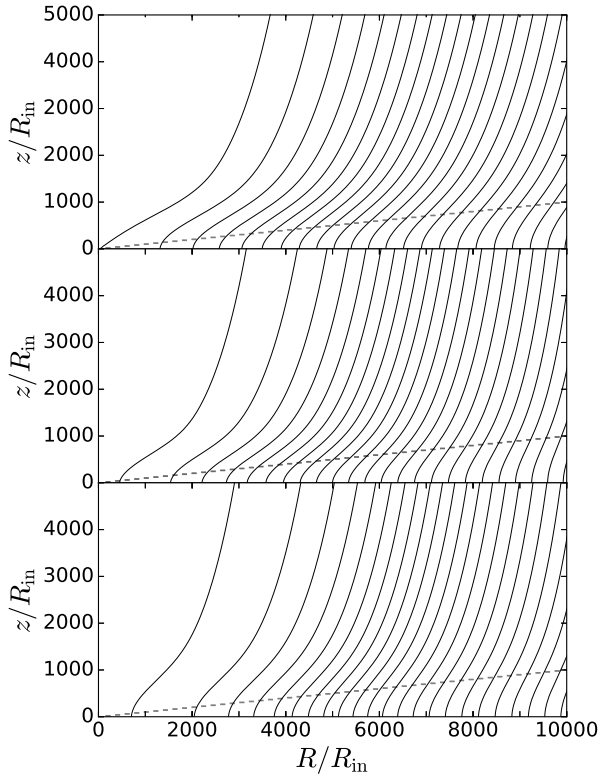}
	\centering
	\caption{The poloidal magnetic field configurations of the disks with different external field strengths. The dashed lines indicate the scale height of the disk extending from $R_{\rm in}$ to $R_{\rm out}=10^4R_{\rm in}$. The Prandt number ${\cal P}_{\rm m}=1$, and the density of the gas at the bottom of the outflow $\beta_{\rm s}=1$ are adopted in the calculations. The results calculated with $\beta_{\rm out}=50, 100, {200}$ are plotted in three panels (from up to down) respectively. Each field line corresponds to a fixed value of $R\psi=\text{const}$,
and the interval between two neighboring field lines is $\Delta R\psi(R,0)=0.058R_{\rm out}\psi_\infty(R_{\rm out}, 0)$, and $ \psi_\infty(R_{\rm out})= B_{\rm{ext}}R_{\rm out}/2$.}
	\label{fig:B_line}
\end{figure}

\begin{figure}
	\centering
	\includegraphics[width=\columnwidth]{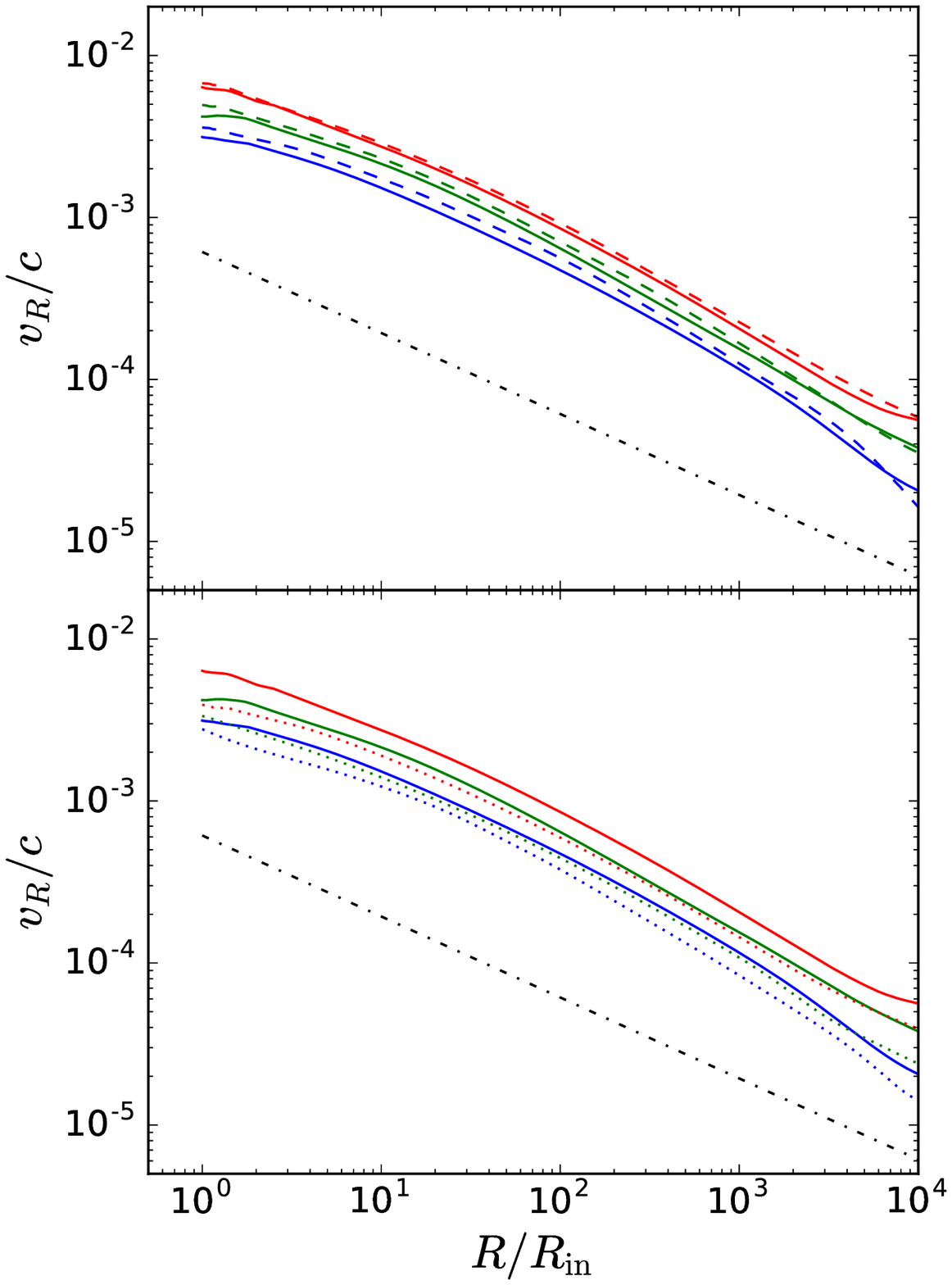}
	\caption{The radial velocity of the accretion disk varies with radius. The color lines indicate the results with different external field strengths, $\beta_{\rm out}= 50 \text{(red)}$, $100 \text{(green)}$, and ${200} \text{(blue)}$. The solid lines correspond to the cases calculated with $\beta_{\rm s }=1$ and ${\cal P}_{\rm m}=1$, {while the dot-dashed lines are the case of a standard thin accretion disk without outflow, as done in \cite{1994MNRAS.267..235L}}. The dashed lines in the upper panel correspond to the results calculated with $\beta_{\rm s }=0.2$, and ${\cal P}_{\rm m}=1$, while the dotted lines in the lower panel are for the results calculated with $\beta_{\rm s }=1$, and ${\cal P}_{\rm m}=1.5$.}
	\label{fig:v_r}
\end{figure}

\begin{figure}
	\centering
	\includegraphics[width=\columnwidth]{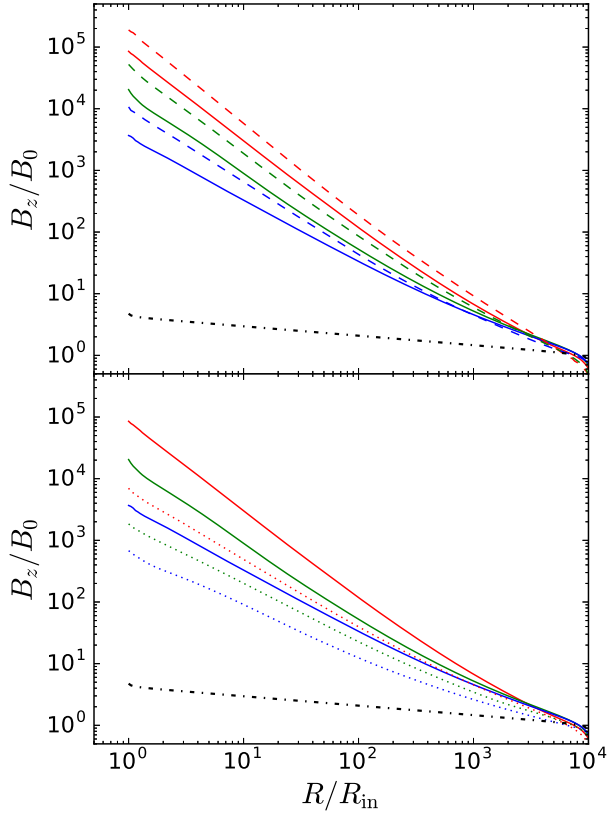}
	\caption{The vertical component of the  poloidal magnetic field varies with radius. The color lines indicate the results with different external field strengths, $\beta_{\rm out}= 50 \text{(red)}$, $100 \text{(green)}$, and ${200}\text{(blue)}$. The solid lines correspond to the cases calculated with $\beta_{\rm s }=1$ and ${\cal P}_{\rm m}=1$, while the dot-dashed lines are the case of a standard thin accretion disk without outflow, as done in \cite{1994MNRAS.267..235L}. The dashed lines in the upper panel correspond to the results calculated with $\beta_{\rm s }=0.2$, and ${\cal P}_{\rm m}=1$, while the dotted lines in the lower panel are for the results calculated with $\beta_{\rm s }=1$, and ${\cal P}_{\rm m}=1.5$.}
	\label{fig:B_field}
\end{figure}

\begin{figure}
	\centering
	\includegraphics[width=\columnwidth]{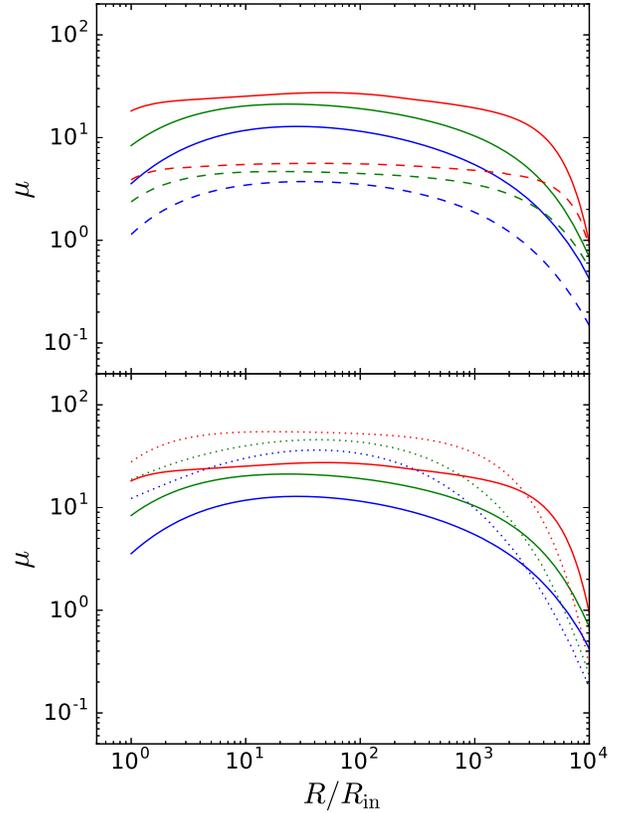}
	\caption{The mass loading parameter $\mu$ varies with radius. The color lines indicate the results with different external field strengths, $\beta_{\rm out}= 50 \text{(red)}$, $100 \text{(green)}$, and ${200} \text{(blue)}$. The solid lines correspond to the cases calculated with $\beta_{\rm s }=1$ and ${\cal P}_{\rm m}=1$. The dashed lines in the upper panel correspond to the results calculated with $\beta_{\rm s }=0.2$, and ${\cal P}_{\rm m}=1$, while the dotted lines in the lower panel are for the results calculated with $\beta_{\rm s }=1$, and ${\cal P}_{\rm m}=1.5$.}
	\label{fig:mu}
\end{figure}

\begin{figure}
	\centering
	\includegraphics[width=\columnwidth]{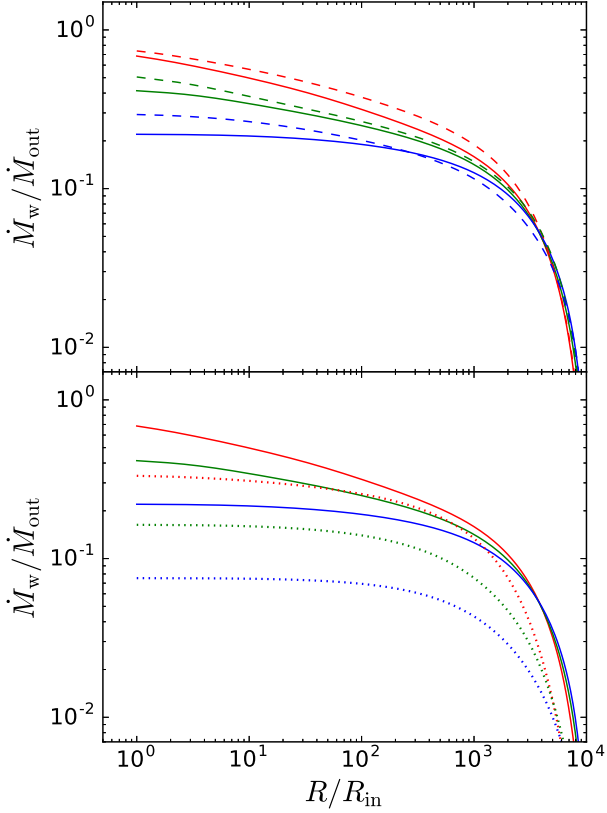}
	\caption{The mass loss rates $\dot{M}_{\rm w}(R)$ in the outflows as functions of radius (integrated from the outer radius of the disk to $R$) with different parameter values. The color lines indicate the results with different external field strengths, $\beta_{\rm out}= 50 \text{(red)}$, $100 \text{(green)}$, and ${200} \text{(blue)}$. The solid lines correspond to the cases calculated with $\beta_{\rm s }=1$ and ${\cal P}_{\rm m}=1$. The dashed lines in the upper panel correspond to the results calculated with $\beta_{\rm s }=0.2$, and ${\cal P}_{\rm m}=1$, while the dotted lines in the lower panel are for the results calculated with $\beta_{\rm s }=1$, and ${\cal P}_{\rm m}=1.5$. The mass accretion rate at the outer edge of the disk is $\dot{M}_{\rm{out}}$.}
	\label{fig:outflow_rate}
\end{figure}

\begin{figure}
	\centering
	\includegraphics[width=\columnwidth]{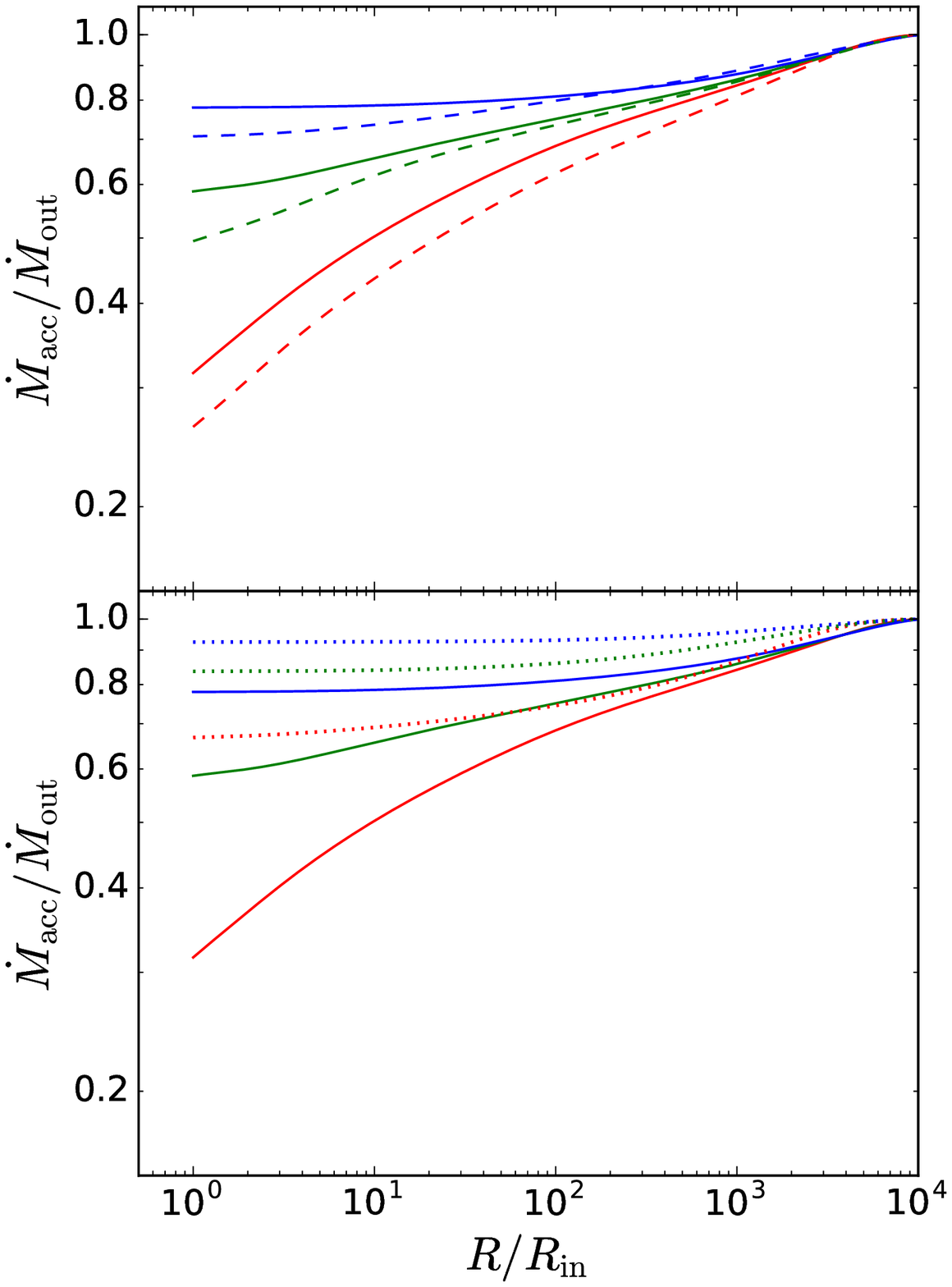}
	\caption{The mass accretion rates $\dot{M}_{\rm acc}(R)$ as functions of radius with different parameter values. The color lines indicate the results with different external field strengths, $\beta_{\rm out}= 50 \text{(red)}$, $100 \text{(green)}$, and ${200} \text{(blue)}$. The solid lines correspond to the cases calculated with $\beta_{\rm s }=1$ and ${\cal P}_{\rm m}=1$. The dashed lines in the upper panel correspond to the results calculated with $\beta_{\rm s }=0.2$, and ${\cal P}_{\rm m}=1$, while the dotted lines in the lower panel are for the results calculated with $\beta_{\rm s }=1$, and ${\cal P}_{\rm m}=1.5$. The mass accretion rate at the outer edge of the disk is $\dot{M}_{\rm{out}}$.}
	\label{fig:acc_rate}
\end{figure}

\begin{figure}
	\includegraphics[width=\columnwidth]{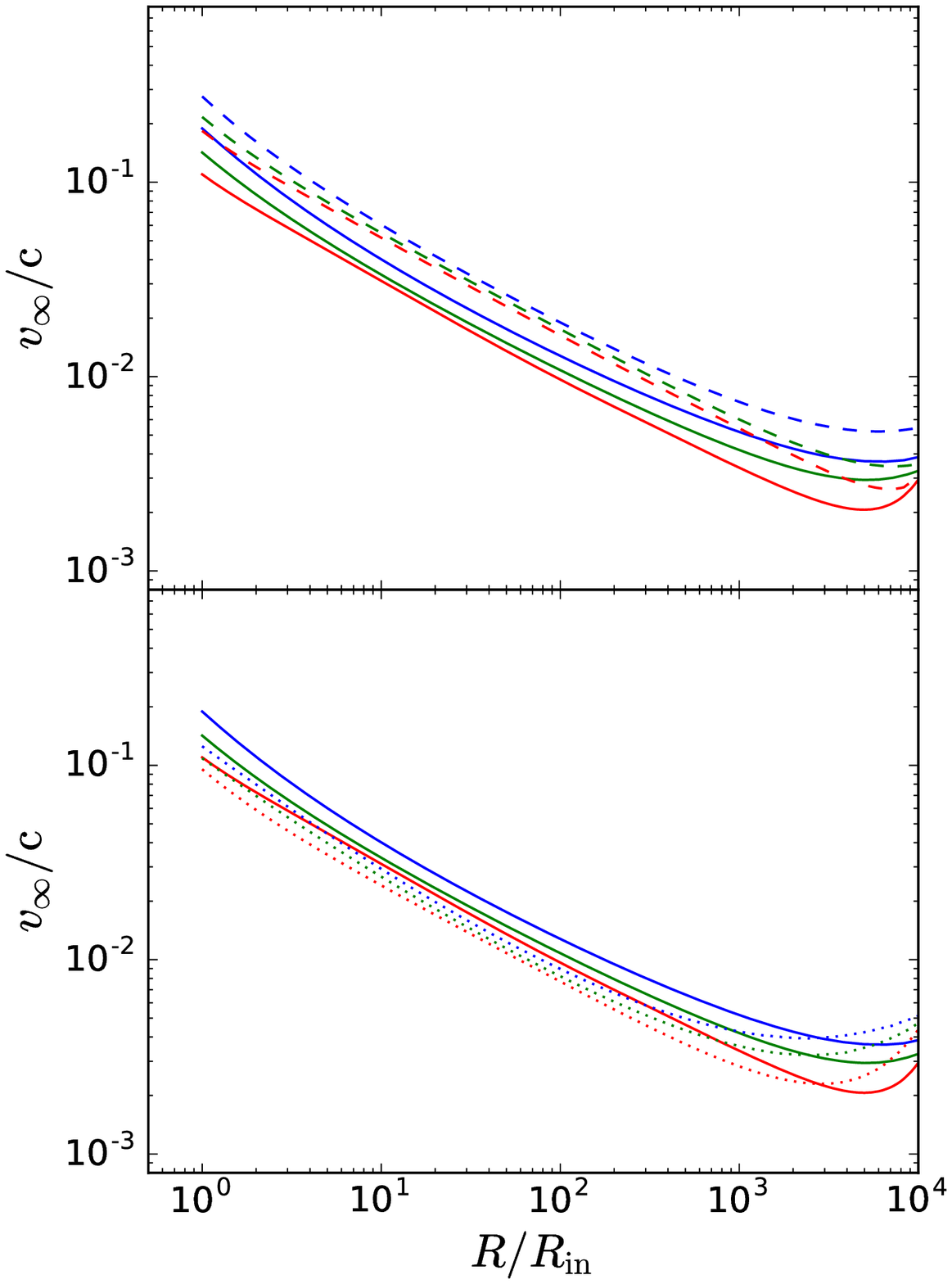}
	\centering
	\caption{The terminal speeds of the outflows as functions of radius with different parameter values. The color lines indicate the results with different external field strengths, $\beta_{\rm out}= 50 \text{(red)}$, $100 \text{(green)}$, and ${200} \text{(blue)}$. The solid lines correspond to the cases calculated with $\beta_{\rm s }=1$ and ${\cal P}_{\rm m}=1$. The dashed lines in the upper panel correspond to the results calculated with $\beta_{\rm s }=0.2$, and ${\cal P}_{\rm m}=1$, while the dotted lines in the lower panel are for the results calculated with $\beta_{\rm s }=1$, and ${\cal P}_{\rm m}=1.5$.}
	\label{fig:v_terminal}
\end{figure}

\begin{figure}
	\includegraphics[width=\columnwidth]{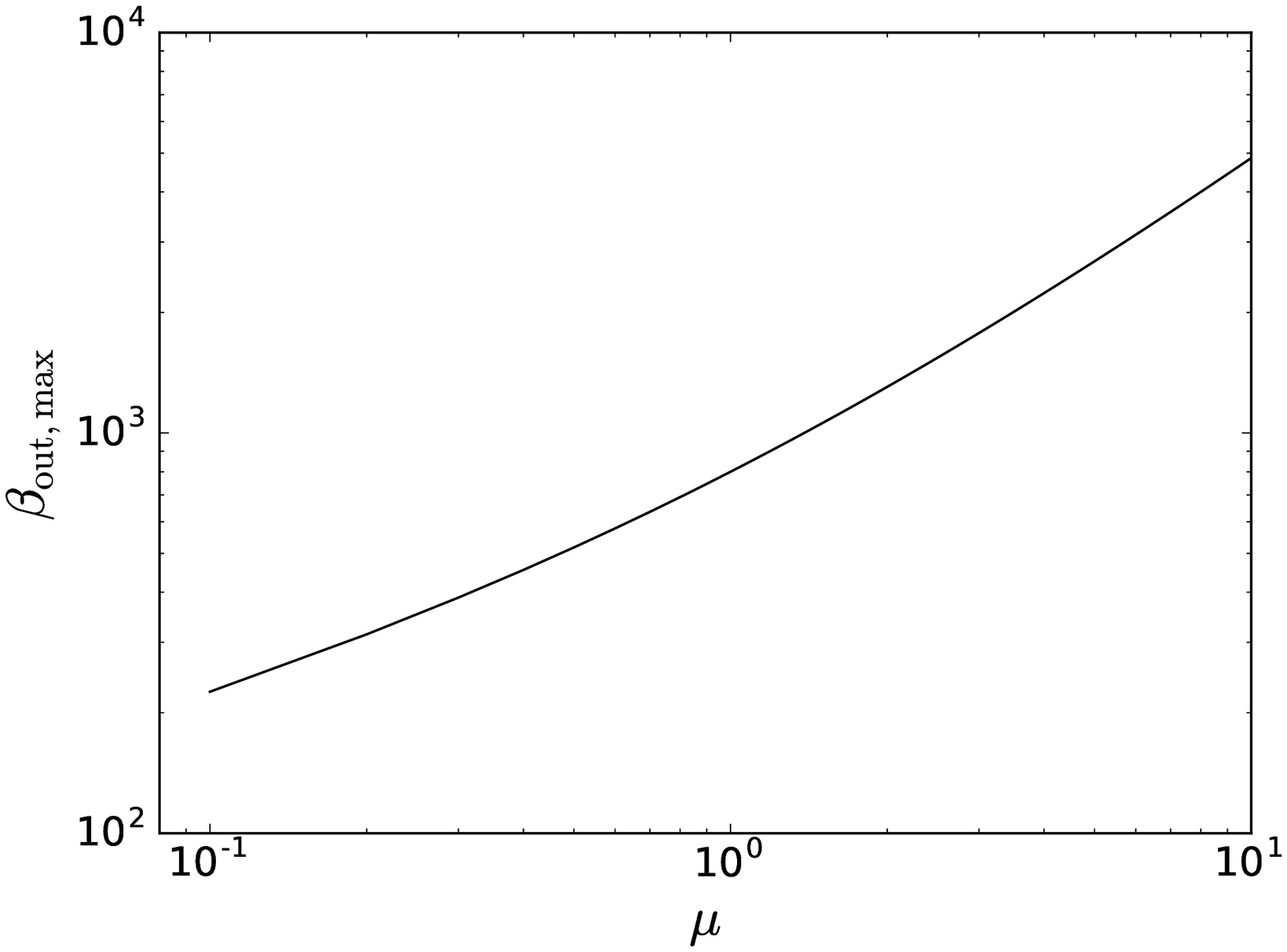}
	\centering
	\caption{The minimal external field strength $\beta_{\rm out,max}$ required at the outer edge of the disk varies with the mass load parameter $\mu$ (see Equation \ref{beta_max}). Typical values of parameters, $\alpha=0.1$ and $H/R =0.1$, are adopted in the estimate. }
	\label{fig:beta_max}
\end{figure}

\section{results and discussion}
\label{result}

    As we know that the magnetic field advected in the accretion disk is a global problem, the
	field strength and the inclination with respect to the disk surface is, in principle, determined
	by the structure of the whole disk \citep*[][]{1994MNRAS.267..235L}.
	We study the global configuration of the large-scale poloidal magnetic field
	advected by the accretion disk with outflows. Most angular momentum of the disk
	is removed by the magnetically driven outflows.
	The disk is coupled with the magnetically driven outflows. Comparing with the previous works, we present self-consistent calculations of an accretion disk-outflow model (see discussion in Section \ref{comparison}). We assume a thin disk with a constant dimensionless temperature $ \Theta =
	c^2_{\rm s,c}/(R^2\Omega^2_{\rm k}) $ independent of radius. For a conventional
	thin disk without magnetic outflows, $ \Theta= (H/R)^2 \simeq\rm {constant} $ is a
	good approximation for a realistic thin disk except the inner region very close to the black hole. A vertical weak external magnetic field
	$ B_{\rm{ext}} $ is assumed to be advected by the disk, which is related to the gas pressure
	of the disk at the outer radius by
	{ $ \beta_{\rm out} = 8 \pi P_{\rm gas}(R_{\rm out})/B^2_{\rm ext}$}.

In this work, the viscosity parameter $\alpha=0.1$ is adopted in all the calculations, which is roughly consistent with that suggested by the numerical simulations \citep*[e.g.,][and the references therein]{2013ApJ...767...30B}. We note that our results are insensitive to the value of $\alpha$.
The temperature of the gas at the mid-plane of the disk $\Theta=0.01$ is adopted in the calculations, which corresponds to a typical thin accretion disk with $H/R\simeq0.1$ in the absence of magnetic field or for the weak field case. The surface temperature of the disk is lower than the temperature at the mid-plane of the disk. It is found that $c_{\rm s,s}=(4/3\tau)^{1/8}c_{\rm s,c}$, which indicates the results depend on the value of $\tau$ rather weakly. We adopt a typical value $\tau=100$ in all the calculations \citep*[][]{2013ApJ...765..149C}. The disk is assumed to extend from $R_{\rm in}$ to $R_{\rm out}=10^4R_{\rm in}$.

The numerical simulations show that the magnetic Prandtl number ${\cal P}_{\rm m}$ is always in a narrow range around unity  \citep*[e.g.,][]{2003A&A...411..321Y,2009A&A...504..309L,2009A&A...507...19F,2009ApJ...697.1901G}. Using the numerical method described in Section. \ref{numerical_method}, we can derive a self-consistent solution of such an accretion disk-outflow system with specified values of the parameters.
The large-scale magnetic field configurations of the disks with outflows are plotted in Figure \ref{fig:B_line} for different external magnetic field strength (i.e., different values of $\beta_{\rm out}$). The magnetic Prandtl number ${\cal P}_{\rm m}=1$ is adopted in the calculations. It is found that the magnetic field is transported inwards and form an open configuration field threading the disk, and the field lines are significantly inclined to the disk, which is a necessary condition for launching outflows from the disk surface \citep*[e.g.,][]{1982MNRAS.199..883B,1994A&A...287...80C,1997MNRAS.291..145C,2000ApJ...538..684P,2012MNRAS.426.2813C}. The results show that the field lines are more inclined to the disk surface for a small $\beta_{\rm out}$, i.e., a strong external field strength case, which implies that a stronger external field is more efficient for field advection in the disk.

The magnetic torque increases with magnetic field strength and field line inclination angle, i.e., the angle of the field line with respect to the disk axis \citep*[e.g.,][]{1994A&A...287...80C,1994MNRAS.268.1010L}. The strength of the magnetic field formed by advection increases with the external field strength, and therefore the magnetic torque due to outflows exerted on the disk increases, which leads to higher radial velocities of the disk. It makes field lines inclined much to the disk surface for a stronger external field (i.e., low-$\beta_{\rm out}$). It is well known that the acceleration of outflows is sensitive to the field line inclination at the disk surface \citep*[][]{1982MNRAS.199..883B}. The field lines with large inclination angle can accelerate outflows more efficiently \citep*[e.g.,][]{1994A&A...287...80C}, which also enhances the magnetic torque.

In Figure \ref{fig:v_r}, we plot the radial velocities of the accretion disks as functions of radius with different values of model parameters. It is found that the radial velocity of the disk with outflows is significantly higher than that of a conventional viscous disk. The strengths of the poloidal magnetic field dragged inwards by thin disk with outflows as functions of radius are plotted in Figure \ref{fig:B_field} with different values of external field strength $\beta_{\rm out}$. The magnetic field strength is significantly enhanced in the inner region of the disk, while the field advection in the conventional viscous accretion disk without outflow is very inefficient (see Figure \ref{fig:B_field}), because the radial velocity of the accretion disk with outflows is substantially higher than that of a standard thin disk (see Equation \ref{v_R}).
	The amplification of the field in the disk is sensitive to the external field strength. A strong external field (low-$\beta_{\rm out}$) leads to strong outflows, which drives the gas in the disk falls to the central black hole rapidly, and therefore the field is dragged inwards efficiently. The strength of the field in the inner region of the disk is higher for a stronger
	external field (i.e., a lower $\beta_{\rm out}$).

In order to explore how the field advection in the disk is affected by the external field strength and Prandtl number, we also plot the results calculated with  $\beta_{\rm s}=0.2$ and ${\cal P}_{\rm m}=1.5$ for comparison. We find that the external field can be efficiently dragged inwards with the field strength amplified several orders of magnitude in the inner region of the disk. The field advection in the standard thin accretion disk without outflow is also plotted in the same figure (see the dot-dashed lines). It is found that the field strength in the inner edge of the disk is only slightly higher than the external field strength, i.e., the field advection is inefficient in a standard thin disk \citep*[][]{1994MNRAS.267..235L}.
We find that the field dragged inwards with outflows is obviously efficient than that in conventional turbulent thin disk. A  fairly strong external field can also be formed if a low Prandtl number ${\cal P}_{\rm m}=1.5$ is adopted, though it is not so strong as that for ${\cal P}_{\rm m}=1$. In this case, more field diffusion in the disk and therefore the field is less enhanced.  

As the gas is magnetically driven from the disk into the outflows, the mass accretion rate is no longer constant in the disk. The mass loading parameter $\mu$ varying with radius with different values of model parameters is plotted in Figure \ref{fig:mu}. The mass loss rates in the outflows integrated from the outer radius of the disk to $R$ are plotted in Figure \ref{fig:outflow_rate}. We plot the mass accretion rates as functions of radius for different values of $\beta_{\rm out}$ in Figure \ref{fig:acc_rate}. The total mass loss rate in the outflows is exactly the difference between the mass accretion rates at the inner and outer edges of the disk, i.e., $\dot{M}_{\rm w}=\dot{M}_{\rm acc}(R_{\rm out})-\dot{M}_{\rm acc}(R_{\rm in})$. It is found that the mass loss rate increases with
{decreasing gas density at} the bottom of the outflow (i.e., $\beta_{\rm s}$), while the mass loss rate decreases with increasing value of ${\cal P}_{\rm m}$. Comparing the results in this figure with those in  Figure \ref{fig:B_field}, we find that much gas is driven into the outflows when the magnetic field is strong.

The terminal speeds of the outflows as functions of radius are plotted for different parameter values in Figure \ref{fig:v_terminal}. It is found that the gas in the inner region of the disk can be accelerated to several ten percent of the light speed, while the terminal speeds of the gas from the outer region of the disk are much lower ($\sim 10^{-3}-10^{-2}~c$). Comparing this figure with Figure \ref{fig:acc_rate}, one finds that a heavy outflow (i.e., with a high mass loss rate) is magnetically accelerated from the disk in the case of a strong external magnetic field (i.e., a small $ \beta_{\rm out}$), while its terminal speed is relatively low.  In this case, the radial velocity is large, so the field is efficiently dragged inwards, the field lines are more inclined to the disk surface (see Figure \ref{fig:B_line}), and the point of the maximum effective potential $ \Psi_{\rm{eff}} $ along the field line is
	close to the disk. The gas in the disk can be relatively easier to overcome such a shallower potential barrier, which leads to a large $ \mu $ {{(see Figure \ref{fig:mu}).} }
	A high-$ \mu $ outflow usually means a heavy or dense outflow, which is too heavy to be accelerated to a high speed by the magnetic field {{(see the upper panel in Figure \ref{fig:B_field} and Figure \ref{fig:mu})} }. The
	terminal velocity of such an outflow is indeed low (see Figures  \ref{fig:acc_rate} and \ref{fig:v_terminal}).	
 The outflow can be accelerated up to a relatively high speed with {$ v_\infty \sim 0.3c$} .
	(see Figure \ref{fig:v_terminal}), which is roughly consistent with the ultra-fast outflows (UFOs)
	detected in some luminous quasars \citep*[][]{2010A&A...521A..57T,2011ApJ...742...44T,2013MNRAS.430...60G}.
	Recent works on outflows in active galactic nuclei (AGN) \citep*[e.g.][]{2015ApJ...805...17F,2018ApJ...853...40F,2018ApJ...852...35K} suggested that the magnetically accelerated outflow is a physically plausible approach for the outflow features observed in X-ray spectroscopic observations.

It was suggested that the highly ionized disk winds/outflows are predominantly driven by magnetic processes (i.e., magnetic pressure and/or magnetocentrifugal), and they are unlikely to be driven by
 the thermal or radiation force, because of their prominently large critical launching radius of winds/outflows and low ultraviolet opacity in the highly ionized winds/outflows \citep*[][]{2006Natur.441..953M,2016ApJ...821L...9M}.
 The outflow seems to have layer-like structure radially, {i.e., the outflow have a velocity gradient along the radius of the disk surface where it is launched (see Figure \ref{fig:v_terminal}, the terminal speeds of the outflows decrease towards the outer edge of the disk). Longterm study of the luminous Seyfert galaxy PG 1211+143 has found multiple blueshifted absorption lines (highly ionized iron) in X-ray spectra with outflow velocities of $ \sim 0.06c,  0.13c,$ and $0.18c $ respectively, which implies that the outflows are not moving outwards at a single velocity, but are likely to have some layer-like structure with multiple velocities} \citep*[][]{2003MNRAS.345..705P,2016MNRAS.457.2951P,2016MNRAS.459.4389P}. The terminal speeds of the gas in the outflow driven from the inner edge of the disk can be as high as $\sim 0.1-0.3~c$, while they decrease with increasing radius, and they are around $10^{-3}-10^{-2}~c$ for the gas from the outer region of the disk (see Figure \ref{fig:v_terminal}),  which is roughly consistent with the ultra-fast outflows (UFOs) detected in some luminous quasars. Blueshifted absorption lines in the X-ray spectra with a typical velocities of several percent of light speed are a common feature in luminous AGNs \citep*[][]{2011ApJ...742...44T,2010A&A...521A..57T,2013MNRAS.430...60G}.

The calculations of the outflow solutions are carried out with suitable boundary conditions (i.e., the temperature and density of the gas at the disk surface). In our work, we adopt a typical vertical optical depth of the disk $ \tau=100 $
	to derive the surface disk temperature. We note that the surface temperature of the disk depends weakly with the optical depth (see Equation \ref{c_sc-c_ss}),
	which means the final results may not be altered much if the optical depth is calculated with a more realistic disk model. The density of gas at the
	disk surface is limited by the gas pressure, which should be lower than the magnetic pressure at the bottom of the outflow (i.e., $ \beta_{\rm s } \lesssim 1 $). We calculate the cases with a lower gas density at the bottom of the outflow ($ \beta_{\rm s }=0.2 $).   It is found that the outflows can be accelerated to a higher terminal velocity for a lower density (see Figure \ref{fig:v_terminal}).

We note that the external field strength $\beta_{\rm out}$ is an important parameter in our model calculations. It is obvious that there is a certain external field strength, below which the accretion disk-outflow system cannot be maintained. As discussed in Section \ref{model}, a strong field of the thin disk can be formed only if the angular momentum of the gas in the disk is removed predominantly via magnetic outflows. Thus, we can roughly estimate the minimal external field strength (i.e., $\beta_{\rm out,max}$) by assuming $ v_{\rm{R,m}} \sim v_{\rm{R,vis}}$ in Equation (\ref{v_R}),
\begin{equation}\label{beta_max}
\beta_{\rm{out}} \la \beta_{\rm out,max}\sim 4 \alpha ^{-1} \mu \left( 1+\mu^{-2/3}\right) \left(\frac{H}{R} \right)^{-1},
\end{equation}
where the approximations $ B_{\rm{z}}\sim B_{\rm p} $ and $ \Omega \sim \Omega_{\rm k}$ are adopted. We plot $\beta_{\rm out,max}$ as a function of $\mu$ in Figure \ref{fig:beta_max}. We find that $\beta_{\rm out}$ is required to be lower than several hundred in order to form a strong field in a thin disk with outflows.

In our model calculations, only poloidal field is considered. However, radial field component will be sheared into azimuthal field component by differential rotation of the disk, which will trigger magnetorotational instability (MRI) \citep*[][]{1991ApJ...376..214B}. \cite{2018MNRAS.473.2791D} performed a global study on the MRI in an accretion flow with differential rotation. They found that a very strong toroidal field in the accretion flow highly suppresses the MRI-dynamo process. \cite{2016MNRAS.460.3488S} also found that the strong toroidal magnetic field generated in their simulation. The MRI enhances the turbulence viscosity, which is
	important in understanding of how turbulence arises and transports angular momentum in astrophysical accretion disks \citep*[][]{1991ApJ...376..214B,1998RvMP...70....1B}. It was found by \citet{2013ApJ...767...30B,2013ApJ...769...76B} that the increasing of vertical field strength can suppress the MRI and a strong magnetocentrifugal wind is launched. In summary, such toroidal field is indeed present in the disk, and it plays an important role in the angular momentum transfer of the gas in the disk mainly through triggering turbulence, though the detailed physics still needs further investigations most probably by numerical simulations. Thus, the $\alpha$-viscosity can still describe the general features of angular momentum transfer due to turbulence in the disk fairly well even if the toroidal field is properly considered. In our model calculations, most angular momentum of the gas in the disk is removed by the outflows, which dominates over that by turbulence in the disk. This implies that our main conclusions will not be altered even if the toroidal field in the disk is considered. We note that the outflows may also be magnetically driven from the hot gas (corona) above the disk \citep*{2014ApJ...783...51C}, and there is indeed observational evidence that hot plasma may help launching jets in X-ray binary and AGNs \citep*[][]{2011MNRAS.416.1324Z,2013ApJ...770...31W}, which is beyond the scope of this work.

\section{SUMMARY}

In this work, a self-consistent global accretion disk-outflow model is constructed, in which the large-scale magnetic field is formed by the advection of the external weak field in the disk. This field is co-rotating with the accretion disk, which accelerates a fraction of the gas into outflows. The disk is driven both by turbulence and the angular momentum loss in magnetic outflows.

We find that a strong large-scale magnetic field can be formed even if an external magnetic field with moderate strength is present ($\beta_{\rm out}$ can be hundreds at the outer edge of the disk). The field lines are inclined substantially to the disk surface, which help launching outflows. The amplification of the field in the inner region of the disk increases with the external field strength. Low gas density at the bottom of the outflows leads to strong outflows with large mass loss rate and high terminal speeds, and then strong magnetic torque, which leads to strong field enhancement in the disk.

The outflow seems to have layer-like structure radially. The terminal speed of the gas in the outflow varies from $\sim 10^{-3}-10^{-2}~c$ to $\sim 0.1-0.3~c$, depending on the location of the gas driven from the disk, which is roughly consistent with the UFOs detected in some luminous quasars.The further detailed comparison of the density and velocity structure of the outflow with the observations will help understanding the physics of outflow.

\section*{Acknowledgements}
We thank the referee for the helpful comments/suggestions that improved the presentation of the manuscript substantially. This work is supported by the NSFC (grants 11773050 and 11833007), the CAS grant (QYZDJ-SSW-SYS023).

\end{document}